\documentclass[
  english,
  aps,
  prl,
  reprint,
  floatfix,
  footinbib,
  preprintnumbers,
  superscriptaddress,
  longbibliography
]{revtex4-2}

\usepackage[normalem]{ulem}
\usepackage{mathtools}
\usepackage{amsmath, amssymb, graphicx, color, hyperref}
\usepackage{float}
\usepackage[english]{babel}
\usepackage{mathrsfs}
\allowdisplaybreaks[3]
\def\dd{\mathrm{d}}
\def\i{\mathrm{i}}
\newcommand{\e}{\mathrm{e}}
\renewcommand{\Im}{\operatorname*{Im}}
\renewcommand{\Re}{\operatorname*{Re}}

\begin{document}

\title{Analytic Bootstrap of the Veneziano Amplitude}

\preprint{USTC-ICTS/PCFT-26-28}

\author{Shi-Lin Wan}
\email{wsl9868@mail.ustc.edu.cn}
\affiliation{Interdisciplinary Center for Theoretical Study, University of Science and Technology of China, Hefei, Anhui 230026, China}%

\author{Shuang-Yong Zhou}
\email{zhoushy@ustc.edu.cn}
\affiliation{Interdisciplinary Center for Theoretical Study, University of
Science and Technology of China, Hefei, Anhui 230026, China}
\affiliation{Peng Huanwu Center for Fundamental Theory, Hefei, Anhui 230026, China}

\date{\today}%

\begin{abstract}
We analytically prove that the Veneziano amplitude is the unique outcome of a dual bootstrap based on dispersive sum rules, unitarity, and some additional stringy input.
This stringy input can be either
the string monodromy condition or the recently uncovered splitting and hidden-zero conditions.
A key ingredient in our proofs is to interpret the
dispersive sum rules as sequences of moments. Also important is the precise incorporation of the extra stringy input into the amplitude ansatz.
Together, these ingredients make the bootstrap analytically tractable and uniquely fix the Veneziano amplitude.
\end{abstract}

\maketitle

The modern S-matrix bootstrap has revived an old idea in a sharper form: basic principles of scattering such as analyticity, crossing symmetry and unitarity, can be surprisingly constraining
\cite{Adams:2006sv,Paulos:2016but,Paulos:2017fhb,deRham:2017avq,Karateev:2019ymz,Remmen:2020vts,Zhang:2020jyn,Huang:2020nqy,Bellazzini:2020cot,Tolley:2020gtv,Caron-Huot:2020cmc,Arkani-Hamed:2020blm,Sinha:2020win,Li:2021lpe,Guerrieri:2021tak,Bern:2021ppb,Chiang:2021ziz,Guerrieri:2021ivu,Caron-Huot:2021rmr,Caron-Huot:2022ugt,Chiang:2022ltp,Henriksson:2022oeu,EliasMiro:2022xaa,Fernandez:2022kzi,Ma:2023vgc,He:2023lyy,Chiang:2023quf,Haring:2023zwu,Berman:2023jys,Berman:2024wyt,Wan:2024eto,Cheung:2025nhw,Berman:2025owb,deRham:2025vaq,Chang:2026ztn,Calisto:2026pvv}
(see \cite{Kruczenski:2022lot,deRham:2022hpx} for a concise review).
One of the main ways to implement these principles, especially when focusing on the positivity part of unitarity, is through dispersive sum rules, which are intimately connected to the mathematical moment problem \cite{Arkani-Hamed:2020blm,Bellazzini:2020cot,Chiang:2021ziz,Chiang:2022ltp,Wan:2024eto,Calisto:2026pvv,schmudgen2017moment}.
The moment approach provides an analytic handle on positivity bounds, and can also allow one to reconstruct the UV theory from IR information, sometimes uniquely \cite{Wan:2024eto, Calisto:2026pvv}.
String theory is a striking arena in which such rigidity can be probed.
In the bootstrap approach, we may also ask: how much of the string amplitude is already forced once some stringy S-matrix data is added to analyticity and positivity?

The simplest testing ground is the 4-point open-string amplitude.
After stripping color and kinematic factors, it is famously given by the Veneziano amplitude
    \begin{align}
        A_{\rm V}(s,t) = - \frac{\Gamma(-s)\Gamma(-t)}{\Gamma(1-s-t)},
    \end{align}
which occupies a special place in the history of the S-matrix \cite{Veneziano:1968yb}.
(We have chosen units such that $\alpha'=1$.) Although many well-motivated deformations have been explored
\cite{Caron-Huot:2016icg,Sever:2017ylk,Guerrieri:2021ivu,Arkani-Hamed:2022gsa,Figueroa:2022onw,Huang:2022mdb,Maldacena:2022ckr,Cheung:2022mkw,Bhardwaj:2022lbz,Chen:2023dcx,Cheung:2023adk,Jepsen:2023sia,Cheung:2023uwn,Geiser:2023qqq,Arkani-Hamed:2023jwn,Bhardwaj:2024klc,Arkani-Hamed:2024nzc,Gadde:2025fil,Cheung:2025nhw,Bhat:2025zex}, recent numerical bootstrap studies provide renewed evidence that its distinctive structure is not accidental.
It has been shown that positivity together with the string monodromy condition can constrain generic lowest-order Wilson coefficients to tiny islands/strips around the open-string point \cite{Huang:2020nqy,Chiang:2023quf,Berman:2023jys}, while the higher-point splitting and hidden-zero conditions can also place strong bounds on the lowest-order coefficients \cite{Berman:2025owb}.
The rigidity of the Veneziano amplitude can be alternatively seen from other perspectives such as UV softness and level truncation/minimal zeros, minimal entanglement or maximal supersymmetry \cite{Cheung:2024uhn, Cheung:2025tbr, Bhat:2024agd, Elvang:2026pmc}
(see \cite{Geiser:2022exp,Cheung:2024obl,Cheung:2025tbr,Berman:2026ezk} for the rigidity of the Virasoro-Shapiro amplitude).
These results point to a concrete possibility: the Veneziano amplitude may be an isolated solution of a stringy S-matrix bootstrap.

In this letter, we prove this expectation analytically.
We start with an amplitude ansatz of the form
\begin{align}
    A(s,t) = -\frac{1}{st} +\!\sum_{p\ge q\ge0} \!a_{p,q}s^{p-q}t^q ,
    ~~ 0<|s|,|t|<\Lambda^2 ,
    \label{eq:EFTpara}
\end{align}
where the EFT cutoff $\Lambda$ is chosen to be below the first massive state. By using the standard positivity bound method and additionally imposing {\it either} the string monodromy condition {\it or} the splitting and hidden-zero conditions, we will rigorously prove that the coefficients $a_{p,q}$ are exactly the Veneziano ones
\begin{align}
    a^{\rm V}_{p,q}=\zeta(p-q+2,\{1\}^{q}) ,
\end{align}
where $\zeta(p_1,\ldots,p_d)$ are multiple zeta values (MZVs; see \hyperref[sec:MZVs]{Appendix~A}). Although the two proofs proceed somewhat differently, in both cases the moment-problem approach to positivity bounds plays a pivotal role. The dispersive sum rules imply that, at each fixed $q$, the sequence of coefficients $a_{p,q}$ forms a Hausdorff moment sequence.
This converts dispersive positivity, together with either stringy input, into a powerful uniqueness statement.

Despite their rather different appearances, the two stringy inputs can be cast into the same $2$-cocycle equation with additive kinematic arguments \footnote{In group cohomology, a multiplicative $2$-cocycle on a group $G$, with trivial $G$-action on the coefficients, is a function $\omega:G\times G\to\mathbb{C} \setminus \{0\}$ satisfying $
(\delta_2\omega)(g_1,g_2,g_3) \equiv {\omega(g_2,g_3)\omega(g_1,g_2g_3)}/({\omega(g_1g_2,g_3)\omega(g_1,g_2)}) = 1$. This is the analogue of the closedness condition $\dd \varphi=0$ in de Rham cohomology.
},
\begin{align}
B(s,u)B(s+u,t)=B(u,t)B(s,u+t),
\label{eq:cocycle_equation}
\end{align}
where $B(s,t)\!\equiv\! -stA(s,t)$. The splitting conditions~yield this equation for independent $s$, $t$ and $u$, while the monodromy condition is equivalent to its restriction to $s+t+u=0$, supplemented by $B(s,-s)=\pi s/\sin(\pi s)$. (In the worldsheet description, this common structure can be understood as arising from the consistency between different decompositions of the same integral containing the Koba--Nielsen factor, either across different branches or at special factorization points \cite{Plahte:1970wy,Bjerrum-Bohr:2009ulz,Stieberger:2009hq,Cao:2024gln}.) The solutions to these cocycle problems then provide another key ingredient in completing the analytical proof.

In what follows, we first formulate the moment approach based on dispersion relations, and then bootstrap the Veneziano amplitude in two separate ways, using as additional input either the string monodromy condition or the splitting and hidden-zero conditions.

\noindent{\bf Sum rules as moments}
As usual in the dual bootstrap approach, we make use of the dispersion relations. We are interested in a color- and kinematically stripped two-channel amplitude $A(s,t)$ that obeys an unsubtracted dispersion relation at small fixed $t$:
\begin{align}
    A(s,t) = -\frac{1}{s t}
    + \int_{\Lambda^2}^{\infty} \frac{\dd s'}{\pi} \frac{\operatorname{Im} A(s',t)}{s'-s},
    \label{eq:dispersion}
\end{align}
where $\Lambda$ is taken to be smaller than the mass of the first heavy state $1/\sqrt{\alpha'}=1$.
The imaginary part of the amplitude admits a $D$-dimensional partial-wave expansion:
$\operatorname{Im} A(s',t) =\! \pi \sum_{\ell=0}^{\infty} G_{\ell}^{(D)} \left(1+\frac{2t}{s'}\right) \rho_{\ell}(s')$,
where we have defined $G_{\ell}^{(D)}(z)= C_{\ell}^{(D-3)/2}(z)/ C_{\ell}^{(D-3)/2}(1)$ and $C_{\ell}^{(D-3)/2}(z)$ is the standard Gegenbauer polynomial.
We will only use the positivity part of partial wave unitarity: $\rho_{\ell}(s')\geq 0$.
Using the EFT ansatz \eqref{eq:EFTpara} and matching to the right-hand side of the dispersion relation, we then get the sum rules
\begin{align}
    \label{eq:sum_rules}
    &(\Lambda^2)^{p} a_{p+q,q} = \int_{0}^{1} x^p \rho_{q} (x) \dd x,~ x\equiv \Lambda^2/s^\prime \in[0,1], \\
    &~~~~\rho_{q} (x) \equiv \sum_{\ell=q}^{\infty} \prod_{k=1}^{q} \frac{
    \mathcal J^2(\ell)-\mathcal J^2(k-1)}{k(k+D/2-2)} \frac{\rho_{\ell}\left({\Lambda^2}/{x}\right)}{(\Lambda^2)^{q} x^{1-q}} ,
    \label{eq:apq_dispersion}
\end{align}
where $\mathcal J^2(\ell)=\ell(\ell+D-3)$. Note that we have written these sum rules as $q$-labeled sequences of Hausdorff moments, thanks to the fact that $\rho_{q} (x)\geq 0$.
\smallskip

\noindent{\bf Veneziano from monodromy:~leading orders} Before proceeding to the full proof that the string monodromy condition can be used in the dual bootstrap to uniquely select the Veneziano amplitude, it is instructive to demonstrate how this analytic bootstrap works at the leading $q$-orders.

In open string scattering, the string monodromy condition reflects the fact that different stripped amplitudes are not independent: analytic continuation of the vertex-operator positions around branch points on the worldsheet relates different orderings with characteristic phase factors \cite{Plahte:1970wy,Bjerrum-Bohr:2009ulz,Stieberger:2009hq}.
Interestingly, it can also emerge in the low-energy EFT setting from locality together with the KKBCJ/double-copy construction \cite{Chen:2022shl}.
In terms of the tree-level stripped amplitude, it can be written simply as
\begin{align}
    \label{eq:sm}
    &\mathcal{S}_{s,t,u} \equiv A(s, t) + \e^{\i \pi s} A(s, u) + \e^{-\i \pi t} A(t,u) = 0,
\end{align}
which will be used as our bootstrap input, in addition to analyticity and unitarity.

To proceed, note that the real and imaginary parts of the $t^0$-order (or $q=0$) monodromy condition are respectively
\begin{align}
    \lim_{t\to 0}\mathcal{S}^{\Re}_{s,t,u}&= 2\sum_{p=0}^{\infty} a_{2p,0}s^{2p}
    -\frac{1- s^2 \cos(\pi s) A(s,-s)}{s^2} = 0 , \nonumber \\
    \lim_{t\to 0}\mathcal{S}^{\Im}_{s,t,u}
    &= \sin(\pi s) A(s,-s)-\frac{\pi}{s}=0.
\end{align}
Combining these two equations and using the fact that $({1-\pi s\cot(\pi s)})/(2 s^2) = \sum_{p=0}^{\infty} \zeta(2p+2)s^{2p}$, we get
\begin{align}
    &(\Lambda^2)^{2p} a_{2p,0}
    = (\Lambda^2)^{2p} \zeta(2p+2)
    = \sum_{j=1}^{\infty}\frac{1}{j^2}\left(\frac{\Lambda^2}{j}\right)^{2p}  ,
    \label{eq:a_evenorder}
\end{align}
where we have multiplied the result by $\Lambda^{4p}$ to connect with the sum rules \eqref{eq:sum_rules}.
However, odd-order terms $(\Lambda^2)^{2p+1} a_{2p+1,0}$ are not fixed by the monodromy condition.
To proceed further, we use the fact that $ (\Lambda^2)^{\gamma} a_{\gamma,0}, \gamma = 0, 1,2, \cdots$ forms a Hausdorff moment sequence
\begin{align}
    \mathfrak{m}^\gamma_0 \equiv (\Lambda^2)^{\gamma} a_{\gamma,0} = \int_{0}^{1} x^\gamma \rho_0(x) \dd{x}.
    \label{eq:m_oddeven}
\end{align}
The key point is that a Hausdorff moment sequence is determinate \cite{schmudgen2017moment}, which allows us to reconstruct the odd moments from the even ones \cite{Wan:2024eto}.
In fact, this reconstruction can be done explicitly:
\begin{align}
    \label{eq:rec_formula_m}
    \mathfrak{m}^{2\gamma+1}_0 = \sum_{k_1=0}^{\infty}\sum_{k_2=0}^{k_1} (-1)^{k_1+k_2} \binom{1/2}{k_1} \binom{k_1}{k_2} \mathfrak{m}^{2\gamma+2k_2}_0 ,
\end{align}
thanks to the fact that for $x\in[0,1]$,
\begin{align}
    x&= \bigl(1-(1-x^2)\bigr)^{1/2} = \sum_{k_1=0}^{\infty} \binom{1/2}{k_1}(-1)^{k_1} (1-x^2)^{k_1} \nonumber \\
    &= \sum_{k_1=0}^{\infty} \sum_{k_2=0}^{k_1} (-1)^{k_1+k_2} \binom{1/2}{k_1} \binom{k_1}{k_2} x^{2k_2} ,\label{eq:rec_formula_x}
\end{align}
where generalized binomials are used.
Plugging Eq.~\eqref{eq:a_evenorder} into Eq.~\eqref{eq:rec_formula_m}, interchanging the summation order and then reusing Eq.~\eqref{eq:rec_formula_x}, we infer
\begin{align}
    \Lambda^{4p+2} a_{2p+1,0} =\!\sum_{j\ge 1}\!\frac{1}{j^2}\!\left(\frac{\Lambda^2}{j}\right)^{\! 2p+1} \!\!\!\!=\Lambda^{4p+2} \zeta(2p+3).
\end{align}
Therefore, for all $p\ge 0$, we have $ a_{p,0} = \zeta(p+2)$.

Next, we establish the uniqueness at the $t^1$-order.
First, notice that the monodromy condition implies $A(s,t)=A(t,s)$, and hence $a_{p,q}=a_{p,p-q}$.
It is straightforward to combine this relation with the $t^1$-order of the string monodromy condition $\lim_{t\to 0} \partial_t \mathcal{S}_{s,t,u} = 0$ to obtain
\begin{align}
\label{eq:a2p1000}
    a_{2p+1,1} &= (p+1)a_{2p+1,0} - \sum_{k=1}^{p}\zeta(2k)\,a_{2p-2k+1,0} \, .
\end{align}
Then, plugging in the previous $t^0$-order result $ a_{p,0} = \zeta(p+2)$ and using a slightly modified Euler's classical reduction formula for height-one MZVs \cite{EulerE477AycockTranslation}, we get $a_{2p+1,1} = \zeta(2p+2,1)$.
Again, roughly speaking, half of the EFT coefficients are already fixed by the monodromy condition.

To obtain the other half, we again appeal to the fact that  $\mathfrak{m}^{\gamma}_1 \equiv (\Lambda^2)^{\gamma+1} a_{\gamma+1,1}$
forms a Hausdorff moment sequence.
Following steps analogous to the $\mathfrak{m}^{\gamma}_0$ case, one can show that
\begin{align}
  a_{2p+2,1}= \zeta (2p+3,1) = \sum_{j=2}^{\infty} \left( \dfrac{1}{j^2} \sum_{j_1=1}^{j-1} \dfrac{1}{j_1} \right) \left( \dfrac{1}{j} \right)^{2p+1} . \nonumber
\end{align}
Therefore, at $t^1$-order, we have $a_{p,1} = \zeta(p+1,1)$.

\smallskip

\noindent{\bf Veneziano from monodromy:~all orders}
The $t^0$- and $t^1$-order results illustrate the interplay between monodromy and moment reconstruction. To organize this mechanism at all orders, we first show that the freedom consistent with monodromy can be parametrized as
\begin{align}
\label{eq:A_CAV}
\! A (s, t) = C (\sigma_2, \sigma_3) A_{\rm V} (s,t), ~~A_{\rm V} = - \frac{\Gamma(-s)\Gamma(-t)}{\Gamma(1-s-t)} ,
\end{align}
where $\sigma_2 = s t + s u + t u$ and $\sigma_3 = s t u$. Subject to $s+t+u=0$, this $stu$-symmetric function $C$ depends on only two independent Mandelstam variables and therefore satisfies a series of null constraints \cite{Tolley:2020gtv,Caron-Huot:2020cmc}. We can use these null constraints to derive triangular recursion relations among EFT coefficients at different orders. Then, combined with dispersive positivity, these recursion relations fix $C=1$ through the Hausdorff moment reconstruction, as illustrated in the previous section.

To derive the parametrization \eqref{eq:A_CAV}, first note that the monodromy condition $\mathcal{S}_{s,t,u}=0$ implies $\mathcal{S}^{\Im}_{s,t,u} = 0$ and $A (s, t) = A (t, s)$. Notice that $\mathcal{S}^{\Im}_{s,t,u} = 0$ leads to
\begin{align}
    \dfrac{B(s,u)}{B(t,u)} = \dfrac{s u A(s,u)}{t u A(t,u)} = \dfrac{\pi s \sin(\pi t)}{\pi t \sin (\pi s)}.
    \label{eq:B/B=sin/sin}
\end{align}
Taking $t \rightarrow 0$, we have $B(s,-s) = \pi s / (\sin \pi s)$. Thus, with $s+t+u=0$ and crossing symmetry, we recognize this as precisely the cocycle equation \eqref{eq:cocycle_equation}.
To resolve the restricted cocycle equation, our strategy is to factor out a particular solution---the Veneziano amplitude---and characterize the remaining freedom. Taking the logarithm of Eq.~\eqref{eq:B/B=sin/sin} for small $s$ and $t$, we obtain
\begin{align}
\widetilde{C}(s,u) - \widetilde{C}(t,u) = \ln \left( \dfrac{B(s,u)}{B(t,u)} \right) + 2 F_+(t) - 2F_+(s) = 0 , \nonumber
\end{align}
where we have defined
\begin{align}
    \label{eq:C_def}
    \widetilde{C} (s,t) &\equiv \ln B (s, t) - F_+(s) - F_+(t) + F_+(u), \\
    F_+(z) & \equiv -\frac{1}{2}\log\frac{\sin(\pi z)}{\pi z} \equiv \sum_{\text{even } k \geq 2} \frac{\zeta(k)}{k}z^{k}.
\end{align}
Clearly, $\widetilde{C}(s, t)$ is analytic around $(s,t) = (0,0)$ and satisfies $\widetilde{C}(0,0) = 0$.
The crossing symmetry $A(s,t)=A(t,s)$ then implies $\widetilde{C}(s,t) = \widetilde{C}(t, s)$. Therefore, $\widetilde{C}(s,t)$ is $stu$-symmetric, and we may rewrite $\widetilde{C} (s, t)=\widetilde{C} (\sigma_2,\sigma_3)$.
Then, Eq.~\eqref{eq:C_def} can be recast as
\begin{align}
    B (s, t) &= \exp \Big(F_+(s)+F_+(t)-F_+(s+t) + \widetilde{C}(\sigma_2,\sigma_3) \Big). \nonumber
\end{align}

Next, notice that the Veneziano amplitude can be written as \cite{Geiser:2022icl}
\begin{align}
    &A_{\mathrm{V}}(s,t) =-\frac{1}{st}\exp\bigl(F_{\rm V}(s)+F_{\rm V}(t)-F_{\rm V}(s+t)\bigr),\\
    &F_{\rm V}(z) \equiv F_+(z)+F_-(z),~ F_-(z) \equiv \sum_{\text{odd } k \geq 3} \frac{\zeta(k)}{k}z^{k} .
\end{align}
Since $F_-(s)+F_-(t)-F_-(s+t) = F_-(s)+F_-(t)+F_-(u)$ is $stu$-symmetric, we can write $\widetilde{C}(\sigma_2,\sigma_3) = F_-(s)+F_-(t)-F_-(s+t)+\hat{C}(\sigma_2,\sigma_3)$.
Thus, we can adopt the following parametrization for $B (s, t)$:
\begin{align}
    \label{eq:B=B_V*C}
    &B (s, t) = B_{\mathrm{V}} (s, t) C (\sigma_2,\sigma_3),~~~B_{\rm V} \equiv - s t A_{\rm V} .
\end{align}
This ansatz fully takes into account the monodromy condition\,\footnote{Strictly speaking, we have only shown in the above that the conditions $\mathcal{S}^{\Im}_{s,t,u} = 0$ and $A (s, t) = A (t, s)$ imply Eq.~\eqref{eq:B=B_V*C}.
However, it is easy to see that Eq.~\eqref{eq:B=B_V*C} implies $\mathcal{S}_{s,t,u} = 0$.
Therefore, the monodromy condition is equivalent to $\mathcal{S}^{\Im}_{s,t,u} = 0$ and $A (s, t) = A (t, s)$.}.

Next, we will derive the triangular recursion relations for EFT coefficients $a_{p,q}$ using the above parametrization and null constraints.
Since $C(\sigma_2,\sigma_3)=e^{\hat{C}(\sigma_2,\sigma_3)}$ and $B(s,0) = B_{\rm V}(s, 0)$, we may expand $C$ as
\begin{align}
    \!\!\!\!\!\!\!C (\sigma_2,\sigma_3) & \!=\! 1 \!- \!s t \sum_{p=0}^{\infty} s^p C_p (t/s), ~C_p (x) \equiv \sum_{q=0}^{p} c_{p,q} x^q. \!\!\! \label{eq:Cpara}
\end{align}
For $C(\sigma_2,\sigma_3)$ to be $stu$-symmetric, $c_{p,q}$ must satisfy a series of null constraints, as will be discussed in \hyperref[sec:monodromy_recursion]{Appendix~B}.

For convenience, let us collect all the $c_{p,q}$ coefficients into a vector
\begin{align}
    c&\equiv \big( c_{0,0} \big)^T \!\!\oplus \big( c_{1,0},c_{1,1} \big)^T \!\!\oplus \! \cdots \! \oplus \big( c_{p,0}, \cdots, c_{p,p} \big)^T \!\!\oplus \cdots, \nonumber
\end{align}
and similarly define vectors for the coefficients $a_{p,q}$, $a^{\mathrm{V}}_{p,q}$ and $\Delta_{p,q} \equiv a_{p,q} - a^{\mathrm{V}}_{p,q}$. Now, Eq.~\eqref{eq:B=B_V*C} can be written as
\begin{align}
    \label{eq:C_def_equiv}
    &1 - C (\sigma_2, \sigma_3) = \sum_{n=0}^{\infty} (1-B_{\mathrm{V}})^n [ B_{\mathrm{V}} - B ] .
\end{align}
Plugging $B (s,t)  = 1 - \sum_{p\geq q\geq 0} a_{p,q} s^{p-q+1} t^{q+1}$ and Eq.~\eqref{eq:Cpara} into Eq.~\eqref{eq:C_def_equiv} and matching the powers of $s$ and $t$, we obtain
\begin{align}
\label{eq:cGDelta}
    c = \sum_{n=0}^{\infty} \mathcal{G}^n \Delta,~~~~ \mathcal{G} \Delta \equiv a^{\mathrm{V}}*\Delta ,
\end{align}
where we have defined the convolution matrix $\mathcal{G}$, with the convolution given by
\begin{align}
    &(f*g)_{p,q} \equiv \sum_{\substack{p_1+p_2=p-2, \, q_1+q_2=q-1\\0\le q_1\le p_1,\;0\le q_2\le p_2}} f_{p_1,q_1} g_{p_2,q_2} .
\end{align}
Thus, our central equation \eqref{eq:cGDelta} is turned into the following relations between the $a_{p,q}$ and $c_{p,q}$ coefficients
\begin{align}
\label{eq:cDeltaM}
    &c_{p,q} = \Delta_{p,q} +\sum_{\substack{0 \leq p' \leq p - 2,\, 0 \leq q' \leq q - 1 \\ 0 \leq q' \leq p', \, q' \geq q-p+p'+1}} \mathcal{M}_{p,q;p',q'} \Delta_{p^\prime,q^\prime} ,
\end{align}
where $\mathcal{M}= \sum_{n=1}^\infty\mathcal{G}^n$.

Our desired recursion relations should involve only $\Delta_{p,q} \equiv a_{p,q} - a^{\mathrm{V}}_{p,q}$, with the $c_{p,q}$ dependence eliminated. To achieve this, we use the null constraints to project out $c_{p,q}$: $\widetilde{\mathcal{T}}^{tu} c=0$. The null projector $\widetilde{\mathcal{T}}^{tu}$ must be chosen judiciously such that the resulting recursion relations are triangular, in the sense that the vanishing of lower-order $\Delta_{p,q}$ recursively implies the vanishing of higher-order ones. The construction involves several technical steps, which we defer to \hyperref[sec:monodromy_recursion]{Appendix~B} due to space limitations. The resulting triangular recursion relation is
\begin{align}
    \label{eq:all_recursion}
    &\Delta_{p,q} = - (\mathcal{N} \Delta)_{p,q} - (\mathcal{M} \Delta)_{p,q} - (\mathcal{N} \mathcal{M} \Delta)_{p,q}
\end{align}
when $p-q$ is even. The explicit form of $\mathcal{N}$ is given in \hyperref[sec:monodromy_recursion]{Appendix~B}, but the following steps rely only on the triangularity of $\mathcal{M}$ and $\mathcal{N}$.

For odd $p-q$, however, no such recursion relation exists. We shall appeal to the Hausdorff moment treatment of the sum rules at each $q$ order to reconstruct the other half of the amplitude coefficients with odd $p-q$, as outlined around Eqs.~\eqref{eq:m_oddeven}-\eqref{eq:rec_formula_x} for the example of $q=0$. For higher $q$, the same reconstruction is possible and follows exactly the same logic because the MZV representation shows that the Veneziano coefficient can be written as
\begin{align}
    \Lambda^{2p} a^{\rm V}_{p+q, q} & =  \sum_{ j > j_1 > \cdots > j_q \ge 1}  \frac{1}{j^2 j_1 \cdots j_q} \left( \dfrac{\Lambda^2}{j} \right)^p .
\end{align}

With all the ingredients explicitly specified, the all-order uniqueness proof can be carried out by mathematical induction.
First, we prove $\Delta_{p,q}=0$ for the lowest $q$-orders (and all $p$), done explicitly in the previous section.
Then, we use Eq.~\eqref{eq:all_recursion} to show $\Delta_{p,q+1}|_{p - (q+1) =\text{even}} = 0$.
Finally, we use the determinacy of the Hausdorff moment representation for the sum rules to infer $\Delta_{p,q+1}|_{p - (q+1)=\text{odd}} = 0$. This completes the proof that the positivity bootstrap, supplemented by the monodromy condition, uniquely gives the Veneziano amplitude $a_{p, q} = a_{p,q}^{\rm V}$.

\smallskip

\noindent{\bf Veneziano from splitting} We now turn to the uniqueness proof based on the splitting and hidden-zero conditions.
These conditions are algebraic relations among amplitudes on special kinematic loci in certain theories \cite{Arkani-Hamed:2023swr, Cao:2024gln, Arkani-Hamed:2024fyd}. For string theory, they can be traced to the structure of the Koba--Nielsen worldsheet integrals \cite{Cao:2024gln}.
In our bootstrap setup, they translate to additional constraints on 4-point amplitude coefficients, some of which are non-convex.
We provide two uniqueness proofs based on these constraints: one in the main text, and another, using the Stieltjes-Perron inversion formula, in \hyperref[sec:splitting_alternative]{Appendix~C}.

We start by requiring the color-ordered 4-point amplitude to have hidden zeros: $A_4(s,t)\bigl|_{s+t=0}=0$. 
This implies that the amplitude has the following form
\begin{align}
    \label{eq:Bsu_Def}
    B(s,t) \equiv \frac{-stA_4(s,t)}{s+t}
    = 1-st\!\!\sum_{p\ge q\ge0}\!\!a_{p,q}s^{p-q}t^q .
\end{align}
Additionally, assume the 5-point amplitude satisfies the splitting condition on the locus
$X_{14}=X_{24}+X_{13}$:
\begin{align}
    &A_5[12345] \big|_{X_{14}=X_{24}+X_{13}}\!\! = A_4(X_{13},X_{25})\,A_4(X_{24},X_{35}) ,
    \nonumber
\end{align}
where the planar variables are defined as $ X_{ij}\equiv-\left(p_i+\cdots+p_{j-1}\right)^2$.
Then, using the cyclic invariance $A_5[12345]=A_5[51234]$ and the crossing symmetry of $A_4(s,t)$, we obtain a standard 2-cocycle equation 
\begin{align}
    B(y_1,y_2) B(y_1+y_2,y_3) = B(y_1,y_2+y_3) B(y_2,y_3),
    \label{eq:BBequalBB}
\end{align}
where we have set $X_{13}=y_2$, $X_{24}=y_1$ and $X_{25}=y_3$.
Note that these are the planar variables of a 5-point amplitude, so $y_1$, $y_2$ and $y_3$ are independent.

To solve Eq.~\eqref{eq:BBequalBB}, we take the logarithm and differentiate with respect to $y_3$ at $y_3=0$, obtaining
$\widetilde B^{(0,1)}(y_1,y_2)
=\widetilde B^{(0,1)}(y_1+y_2,0)
-\widetilde B^{(0,1)}(y_2,0)$,
where $\widetilde B(y_1,y_2)\equiv\ln B(y_1,y_2)$.
Integrating with respect to $y_2$ from $0$ to $t$ and setting $y_1=s$ then yields the standard coboundary solution
\footnote{
A $2$-coboundary is a cocycle of the form $\omega=\delta_1 f \equiv f(g_1)f(g_2)/f(g_1g_2)$ for some $f:G\to\mathbb C \setminus \{0\}$. This is analogous to an exact differential form $\varphi=\dd \lambda$ in de Rham cohomology.
}
\begin{align}
\label{eq:lnBsuFFF}
\ln B(s,t)=F(s)+F(t)-F(s+t),
\end{align}
where $F(y)\equiv-\int_0^y\widetilde B^{(0,1)}(y',0)\,\dd y'$.
Thus, the splitting conditions allow us to determine the non-forward coefficients recursively from the forward coefficients $a_{p,0}$ alone. Thanks to dispersive positivity, the forward coefficients form a Hausdorff moment sequence. We now use its generating function to uniquely bootstrap the Veneziano amplitude. 

To that end, using $B(z,0)=1$, Eq.~\eqref{eq:Bsu_Def} and the $q=0$ moment representation \eqref{eq:sum_rules}, we get the generating function:
\begin{align}
    F'(z) &= \sum_{p=0}^{\infty} a_{p,0}z^{p+1} = z\int_0^1 \frac{\rho_0(x)}{1-zx/\Lambda^2}\dd x. 
\end{align}
Integrating with $F(0)=0$ yields
\begin{align}
    F(z) = \int_{\Lambda^2}^{\infty}\!\left[-\ln\!\left(1-\frac{z}{s'}\right)-\frac{z}{s'}\right]\dd\nu_0(s'),
    \label{eq:FofzRep}
\end{align}
where $s'=\Lambda^2/x$ and $\dd\nu_0(s')\equiv\Lambda^2\rho_0(\Lambda^2/s')\dd s'$. Substituting Eq.~\eqref{eq:FofzRep} into Eq.~\eqref{eq:lnBsuFFF}, the linear terms cancel within the integral, so we obtain
\begin{align}
\label{eq:lnBInt}
    \!\!\!\!\!\ln B(s,t) = \int_{\Lambda^2}^{\infty} \ln \dfrac{s'(s'-s-t)} {(s'-s)(s'-t)} \dd{\nu_0}\!(s') .
\end{align}

For a tree-level amplitude, a generic measure takes the form $\dd{\nu_0} = \sum_n \omega_{n}\delta(s'-s_n)\dd{s'}$.
Without loss of generality, let us label the pole locations as $s_1<s_2<\cdots$. Substituting this measure into Eq.~\eqref{eq:lnBInt} gives
\begin{align}
\label{eq:B_product_form}
    B(s,t)= \prod_n \left( \dfrac{s_n (s_n-s-t)}{(s_n-s)(s_n -t)} \right)^{\omega_n} .
\end{align}
The key observation is that the exponents must be
\begin{align}
    \omega_n = 1 ,
\end{align}
since a tree-level amplitude only has simple poles.

Let us now look at the residue of $A_4(s,t)$ at the first massive pole $s=s_1$:
\begin{align}
\label{eq:R1tForm}
    &R_1(t) = - \dfrac{s_1+t}{s_1-t} \prod_{n \neq 1} \dfrac{s_n(s_n-s_1-t)} {(s_n-s_1)(s_n-t)}.
\end{align}
For a genuine interacting theory with finitely many spins at the first massive pole (as expected from locality \cite{Cheung:2022mkw}), partial-wave unitarity together with the positivity of the Gegenbauer polynomials implies \cite{deRham:2017avq}: $R_1(t>0)<0$.
This means that $R_1(t)$ must be free of zeros for $t>0$. Given Eq.~\eqref{eq:R1tForm}, $R_1(t>0)<0$ then implies that
\begin{align}
    \label{eq:Finalspectrum}
    s_n=ns_1,~~ n=1,2,3,....
\end{align}
To see this, note that for $R_1(t>0)$ to be free of zeros, the apparent zeros at $t=s_n-s_1,~n=2,3,...$ must be canceled by poles at $t=s_m,~m=1,2,...$. 
That is, we must have $s_n-s_1\in \{s_m\}_{m\geq 1}$. From using this condition together with $s_1<s_2<\cdots$, we can readily deduce $s_n-s_1=(n-1)s_1$ by induction.

Substituting the spectrum \eqref{eq:Finalspectrum} into Eq.~\eqref{eq:B_product_form}, we arrive at a well-known product form of the Veneziano amplitude \cite{Geiser:2022icl}.
This completes the proof.
\smallskip

\noindent{\bf Summary} In this letter, we have rigorously proven the uniqueness of the Veneziano amplitude from the standard dual bootstrap assumptions supplemented by additional stringy input motivated by the worldsheet construction.
When this input is the open-string monodromy condition, roughly speaking, monodromy fixes half of the amplitude coefficients recursively to be those of the Veneziano amplitude, and the Hausdorff moment representation then reconstructs the other half.
When this input is the splitting and hidden-zero conditions, the moment method allows us to parametrize the generic amplitude in a convenient way; unitarity of the first massive residue then forces the spectrum to lie on the Regge trajectory. These two inputs share the same underlying cocycle structure, albeit in different forms, and resolving the corresponding equations provides another key component of the analytic proof. Thus, two different simple pieces of stringy input, monodromy and splitting, lead to the same unique answer: the Veneziano amplitude.

Our work highlights the usefulness of the moment-problem approach in analytic S-matrix bootstrap. It would be interesting to investigate whether similar ideas can be applied in other bootstrap settings.
~\\

\noindent{\bf Acknowledgment}
We would like to thank Fu-Ming Chang, Jian Xin Lu and Zhuo-Hui Wang for helpful discussions. SYZ acknowledges support from the National Natural Science Foundation of China under grant No.~12475074 and No.~12247103.

\bibliography{ref}

\clearpage
\section*{Appendices}
\label{sec:appendix}

\section{A. Multiple zeta values}
\label{sec:MZVs}

Multiple zeta values (MZVs) can be defined by the Euler-Zagier sum \cite{hoffman1992multiple, zagier1994values, ZudilinMZVNotes}
\begin{align}
    \zeta(p_1,\ldots,p_d) \equiv \sum_{\infty > j_1 > \cdots > j_d \ge 1} \frac{1}{j_1^{p_1}\cdots j_d^{p_d}},
\end{align}
where $d$ is the depth of $\zeta(p_1,\ldots,p_d)$ and the number of $p_i$ is referred to as its height.
The Veneziano amplitude can be expressed using MZVs
\begin{align}
    A_{\rm V}(s,t) =  - \dfrac{1}{s t} +\sum_{p\geq q \geq 0}\zeta(p-q+2,\{1\}^q) s^{p-q} t^q,
\end{align}
where the usual shorthand is adopted
\begin{align}
    \zeta(p,\{1\}^q) = \zeta(p,\underbrace{1,\cdots,1}_q ) .
\end{align}

When verifying the uniqueness at $t^1$-order, we used Eq.~\eqref{eq:a2p1000}, which comes from Euler's classical reduction formula for height-one MZVs \cite{Euler1776E477, EulerE477AycockTranslation}:
\begin{align}
    2\zeta(n,1) = n\zeta(n+1)
    - \sum_{k=1}^{n-2}\zeta(n-k)\zeta(k+1) .
\end{align}
To see this, we can take $n = 2p + 2$ and get
\begin{align}
    &~~~~\zeta(2p+2,1) \\
    &= \dfrac{(2p+2)}{2} \zeta(2p+3) - \dfrac{1}{2} \sum_{k=1}^{2p} \zeta(2p+2-k)\zeta(k+1) \nonumber \\
    &= (p+1) \zeta(2p+3) - \sum_{k=1}^{p} \zeta(2k) \zeta(2p+3-2k) ~. \nonumber
\end{align}

\section{B. Veneziano from monodromy: null-constraint projection}
\label{sec:monodromy_recursion}

In the main text, we have derived Eq.~\eqref{eq:cDeltaM}, which is a relation between $c_{p,q}$ and $\Delta_{p,q} = a_{p,q} - a_{p,q}^{\rm V}$. Note that $\mathcal{M}$ is strictly triangular, since the sums over $p'$ and $q'$ run only up to $p-2$ and $q-1$ respectively and the higher-order $\mathcal{G}^n\Delta$ terms only make $\mathcal{M}$ more triangular.

To achieve an all-order Veneziano bootstrap, we need a triangular recursion relation for $a_{p,q}$ (or equivalently $\Delta_{p,q}$). 
To this end, we use the null constraints to project out $c_{p,q}$ from Eq.~\eqref{eq:cDeltaM}. That is, we shall now use the fact that $C (\sigma_2,\sigma_3)$ is $stu$-symmetric, which means that there exists a matrix $\mathcal{T}$ such that $\mathcal{T} c = 0$. As is well known, the null-constraint matrix $\mathcal{T}$ can be organized by the $p$ orders  \cite{Tolley:2020gtv,Caron-Huot:2020cmc}:
\begin{align}
    \mathcal{T} = \mathcal{T}_0 \oplus \mathcal{T}_1 \oplus \cdots \oplus \mathcal{T}_p \oplus \cdots ,
\end{align}
where $\mathcal{T}_p$ acts on $c_{p,q}$. 
For each $\mathcal{T}_p$, we can further decompose it into two blocks:
\begin{align}
    \mathcal{T}_p = \Big( \mathcal{T}^{st}_p ,~ \mathcal{T}^{tu}_p \Big)^T ,
\end{align}
corresponding to imposing two independent crossing symmetries: $C(\sigma_2,\sigma_3)=C(\sigma_2,\sigma_3)|_{s\leftrightarrow t}$ and $C(\sigma_2,\sigma_3)=C(\sigma_2,\sigma_3)|_{t\leftrightarrow u}$. We can extract the $\mathcal{T}^{st}_p$ and $\mathcal{T}^{tu}_p$ matrices via the following operators
\begin{align}
    \hat{\mathcal{T}}^{st}_p \Big( C_{p} (x) \Big) &\equiv x C_{p} (x) - x^{p+2} (1/x) C_{p} (1/x)=0, \\
    \hat{\mathcal{T}}^{tu}_p \Big( C_{p} (x) \Big) &\equiv x C_{p} (x) + (1+x) C_{p} (-1-x)=0,
\end{align}
which translate to
\begin{align}
    (\mathcal{T}^{st}_p)_{q_1,q_2} &= \delta_{q_1,q_2} - \delta_{p-q_1,q_2}, \nonumber \\
    (\mathcal{T}^{tu}_p)_{q_1,q_2} &= \delta_{q_1,q_2+1} + (-1)^{q_2} \begin{pmatrix}
        q_2+1 \\
        q_1
    \end{pmatrix} \theta (q_2 - q_1+1),\nonumber
\end{align}
where $\theta(n)=1$ for $n\ge 0$ and vanishes otherwise.

Note that none of the above $\mathcal T$ matrices can be directly used to perform our desired projection, which requires that, after the projection, $c_{p,q}$ are completely eliminated, while the resulting equation gives a triangular recursion relation for $\Delta_{p,q}$. Such a goal can be achieved by the matrix $\widetilde{\mathcal{T}}^{tu} = \oplus_p \widetilde{\mathcal{T}}^{tu}_p$, with
\begin{align}
    \widetilde{\mathcal{T}}^{tu}_p \equiv \dfrac{1}{2} \mathcal{R}_{p} \mathcal{T}^{tu}_p ( \mathcal{I}_p - \mathcal{T}_p^{st} ),
\end{align}
where we have introduced $(\mathcal{I}_p)_{q_1,q_2} \equiv \delta_{q_1,q_2}$, $(\mathcal{R}_p)_{q_1,q_2} \equiv \delta_{p+1-q_1,q_2}$.
It is easy to see that $\mathcal{T} c=0$ implies that $\widetilde{\mathcal{T}}^{tu}$ annihilates $c$: 
\begin{equation}
\widetilde{\mathcal{T}}^{tu} c=0 .
\end{equation}
Explicitly, we have
\begin{align}
    &(\widetilde{\mathcal{T}}^{tu}_p)_{q,q'} = \dfrac{1+(-1)^{p-q}}{2} \delta_{q,q'} + \mathcal{N}_{p,q;p,q'} , \\
    & {\rm with}~ \mathcal{N}_{p,q;p',q'} \equiv \dfrac{(-1)^{p-q'}}{2}\! \begin{pmatrix}
        p-q'+1 \\
        p-q+1
    \end{pmatrix}\! \theta(q-q'-1) \delta_{p p'}.\nonumber
\end{align}
Note that the $\mathcal{N}$ matrix is strictly lower triangular in $q$. Using $\widetilde{\mathcal{T}}^{tu}$ to project out the $c_{p,q}$ coefficients in Eq.~\eqref{eq:cDeltaM} leads to the triangular recursion relation in $q$, Eq.~\eqref{eq:all_recursion}, in the main text.

\section{C. Veneziano from splitting: Alternative proof}
\label{sec:splitting_alternative}

Here we present an alternative proof of the uniqueness of the Veneziano amplitude from the splitting and hidden-zero conditions.  The upshot is that, in this proof, we need only assume that the first massive gap corresponds to a pole, while the discreteness of the remaining spectrum emerges as an outcome of the bootstrap.

Let us start from the representation Eq.~\eqref{eq:lnBInt}
of the amplitude, which already incorporates the dispersive nature of the amplitude coefficients, and assume that the first mass gap is at $s_1$ where the amplitude has a simple pole.
This means that the measure in Eq.~\eqref{eq:lnBInt} can be written as
\begin{align}
    \dd{\nu_0(s')} = \delta(s'-s_1)\dd{s'} + \dd{\nu(s')} ,
\end{align}
where $\nu$ only has support on $[s_c,\infty)$ with $s_c>s_1$. Then,
Eq.~\eqref{eq:lnBInt} becomes
\begin{align}
    B(s,t)
    &= \frac{s_1(s_1-s-t)}{(s_1-s)(s_1-t)} \nonumber \\
    &~~~~\times \exp\left[ \int_{s_c}^{\infty} \ln \frac{s'(s'-s-t)}{(s'-s)(s'-t)} \dd\nu(s') \right].
    \label{eq:B_first_pole_split}
\end{align}
The residue $R_1(t)$ at $s=s_1$ is given by
\begin{align}
    \!\!\!\! -R_1(t) = \frac{s_1\!+\!t}{s_1\!-\!t}
    \exp\!\left[ \int_{s_c}^{\infty}\!\! \ln \frac{s'(s'-s_1-t)}{(s'-s_1)(s'-t)} \dd\nu(s') \! \right],
    \label{eq:R1_integral}
\end{align}
from which we get
\begin{align}
    \dfrac{R'_1(t)}{R_1(t)}
    &= \frac{1}{s_1+t} + \frac{1}{s_1-t} \nonumber \\
    &~~~~~ + \int_{s_c}^{\infty} \left( \frac{1}{s'-t} - \frac{1}{s'-s_1-t} \right) \dd\nu(s').
    \label{eq:L1_integral}
\end{align}
Now, the first massive state is assumed to contain only finitely many spins, so $R_1(t)$ is a finite-degree polynomial. As argued in the main text,
partial-wave positivity also gives $R_1(t)<0$ for $t>0$.
Therefore, $R_1'(t)/R_1(t)$ is free of poles
on the positive real $t$-axis. This fact can be used to prove that the apparent singularities on the right-hand side of Eq.~\eqref{eq:L1_integral} must cancel exactly.

To see this, let us first rewrite Eq.~\eqref{eq:L1_integral} as
\begin{align}
    \frac{R_1'(t)}{R_1(t)} = \frac{1}{s_1+t} + r(t), \quad r(t) \equiv \int_{\mathbb{R}^+} \frac{\dd\eta(s')}{s'-t},
    \label{eq:L1_signed_measure}
\end{align}
where
\begin{align}
    \label{eq:detaDef}
    \dd\eta(s') = \delta (s'-s_1) \dd s' +\dd{\nu (s')} -\dd{\nu(s'+s_1)} .
\end{align}
The second term is the Stieltjes transform of the measure $\eta$ \cite{schmudgen2017moment}, which is widely used in the moment problem literature. An important property of the Stieltjes transform is that the measure can be uniquely determined by the Stieltjes-Perron inversion formula
\begin{align}
    &~~~~~~~~~~~\eta (\{ a \}) = - \lim_{\epsilon\to 0^+} \i \epsilon\, r(a+\i\epsilon), \label{eq:stieltjes_inversion_eta} \\
    &\eta((a,b)) = \lim_{\epsilon\to0^+} \int_a^b \!\! \Im \left[ r(t+\i\epsilon) \right] \frac{\dd t}{\pi}
    - \dfrac{\eta (\{ a \}) + \eta (\{ b \})}{2}, \nonumber
\end{align}
where we choose $(a,b)\subset \mathbb{R}^+$. Now, the key point is that $r(t)=R_1'(t)/R_1(t)-1/(s_1+t)$ has no singularity on $\mathbb{R}^+$. The inversion formula then implies that $\eta(\{a\})=0$ and $\eta((a,b))=0$. Thus, Eq.~\eqref{eq:detaDef} can be re-written as
\begin{align}
    \label{eq:measure_shift_identity}
    \dd{\nu(s'+s_1)} = \delta (s'-s_1) \dd {s'} +\dd{\nu (s')}.
\end{align}
This equation can be solved iteratively for different regions.

To this end, let us first consider $s'\in (0,s_1]$.
In this region, we have $\dd{\nu(s'+s_1)} = \delta (s'-s_1) \dd {s'}$, which can be re-written as $\dd{\nu(s')} = \delta (s'-2s_1) \dd {s'},~s' \in (s_1,2s_1]$.
Next, we consider $s'\in (s_1,2s_1]$.
In this region, we have $\dd{\nu(s'+s_1)} = \dd{\nu(s')}$, which, upon using the previous result, leads to $\dd{\nu(s'+s_1)} = \delta (s'-2s_1) \dd {s'}$.
Shifting $s'$ by $s_1$, we get $\dd{\nu(s')} = \delta (s'-3s_1) \dd {s'},~s' \in (2s_1,3s_1]$.
Obviously, this can be generalized to arbitrarily large $s'$. Combining all the regions, we get
\begin{align}
    \dd\nu_0(s') &= \delta(s'-s_1)\dd{s'} + \dd{\nu(s')} \nonumber \\
    &= \sum_{n=1}^{\infty} \delta(s'-n s_1)\,\dd s' .
\end{align}
Plugging this into Eq.~\eqref{eq:lnBInt} and setting $s_1=1$, we finally arrive at a well-known product form of the Veneziano amplitude \cite{Geiser:2022icl}:
\begin{align}
    A_{\rm V} (s,t) = - \dfrac{1}{s \, t} \prod_{n=1}^{\infty} \dfrac{n^2-n(s+t)}{(n-s)(n-t)} .
\end{align}

\end{document}